\documentclass{article}
\usepackage{arxiv}
\usepackage[utf8]{inputenc} 
\usepackage[T1]{fontenc}    
\usepackage{hyperref}       
\usepackage{url}            
\usepackage{booktabs}       
\usepackage{amsfonts}       
\usepackage{nicefrac}       
\usepackage{microtype}      
\usepackage{lipsum}
\usepackage{graphicx}
\usepackage{doi}
\usepackage{lineno,hyperref}
\usepackage{amsmath}
\usepackage{mathtools}
\usepackage{subfigure}
\usepackage{blindtext}
\usepackage{multirow}
\usepackage{bm}
\usepackage{nicematrix}
\usepackage[inline]{enumitem}

\usepackage{multirow}

\usepackage{siunitx}
\usepackage{nicematrix}

\usepackage{blindtext}
\usepackage{hhline}
\usepackage{epstopdf}
\usepackage{tabularx}

\usepackage{bm}
\usepackage[aboveskip=1pt,labelfont=bf,labelsep=period,justification=centering,singlelinecheck=off]{caption}

\newcolumntype{+}{!{\vrule width 2pt}}
\newcolumntype{b}{X}
\newcolumntype{s}{>{\hsize=.5\hsize}X}

\let\svthefootnote\thefootnote
\newcommand\blankfootnote[1]{%
  \let\thefootnote\relax\footnotetext{#1}%
  \let\thefootnote\svthefootnote%
}

\newlength\savedwidth

\makeatletter
\renewcommand{\@biblabel}[1]{\quad#1.}
\makeatother

\setlist[enumerate,1]{%
  label=\arabic*.,
}
\setlist*[enumerate,1]{%
  label=(\roman*),
}
\graphicspath{ {./images/} }

\setlist[enumerate,1]{%
  label=\arabic*.,
}
\setlist*[enumerate,1]{%
  label=(\roman*),
}
\graphicspath{ {./images/} }

\begin{document}

\vspace*{0.2in}

\begin{flushleft}
{\Large
\textbf\newline{SEH: Size Estimate Hedging for Single-Server Queues}
}
\newline
\\
Maryam Akbari-Moghaddam\textsuperscript{1,*} and 
Douglas G. Down\textsuperscript{1}
\\
\bigskip
\textbf{1} Department of Computing and Software, McMaster University, Hamilton, Ontario L8S 4L7, Canada

\bigskip

* Corresponding author \\
Email: akbarimm@mcmaster.ca

\end{flushleft}

\begin{abstract}
For a single server system, Shortest Remaining Processing Time (SRPT) is an optimal size-based policy. In this paper, we discuss scheduling a single-server system when exact information about the jobs’ processing times is not available. When the SRPT policy uses estimated processing times, the underestimation of large jobs can significantly degrade performance. We propose a simple heuristic, Size Estimate Hedging (SEH), that only uses estimated processing times for scheduling decisions. A job's priority is increased dynamically according to an SRPT rule until it is determined that it is underestimated, at which time the priority is frozen. Numerical results suggest that SEH has desirable performance for estimation error variance that is consistent with what is seen in practice.


\end{abstract}

\section{Introduction} \label{Introduction}

Over\blankfootnote{For a published version of this paper refer to Quantitative Evaluation of Systems: 18th International Conference, \url{https://doi.org/10.1007/978-3-030-85172-9_9}.} the past decades, there has been significant study on the scheduling of jobs in single-server queues. When preemption is allowed and processing times are known to the scheduler, the Shortest Remaining Processing Time (SRPT) policy is optimal in the sense that, regardless of the processing time distribution, it minimizes the number of jobs in the system at each point in time and hence, minimizes the mean sojourn time (MST) \cite{schrage1968letter}, \cite{schrage1966queue}. 
However, scheduling policies such as SRPT are rarely deployed in practical settings. A key disadvantage is that the assumption of knowing the exact job processing times prior to scheduling is not always practical to make. However, it is often possible to estimate the job processing times and use this approximate information for scheduling. The Shortest Estimated Remaining Processing Time (SERPT) policy is a version of SRPT that employs the job processing time estimates as if they were error-free and thus, schedules jobs based on their estimated remaining times. Motivated by the fact that estimates can often be obtained through machine learning techniques, Mitzenmacher \cite{mitzenmacher2019scheduling} studies the potential benefits of using such estimates for simple scheduling policies. For this purpose, a price for misprediction, the ratio between a job’s expected sojourn time using its estimated processing time and the job’s expected sojourn time when the job processing time is known is introduced, and a bound on this price is given. The results in \cite{mitzenmacher2019scheduling} suggest that naïve policies work well, and even a weak predictor can yield significant improvements under policies such as SERPT. However, this insight is only made when the job processing times have relatively low variance. As discussed below, when job processing times have high variance, underestimating even a single very large job can severely affect the smaller jobs' sojourn times. 

The work in \cite{mitzenmacher2019scheduling} has the optimistic viewpoint that it is possible to obtain improved performance by utilizing processing time estimates in a simple manner. The more pessimistic view is that when job processing times are estimated, estimation errors naturally arise, and they can degrade a scheduling policy's performance, if the policy was designed to exploit exact knowledge of job processing times \cite{lu2004size}. The SERPT policy may have poor performance when the job processing times have high variance and large jobs are underestimated. Consider a situation where a job with a processing time of 1000 enters the system and is underestimated by 10\%. The moment the job has been processed for 900 units (its estimated processing time), the server assumes that this job's estimated remaining processing time is zero, and until it completes, the job will block the jobs already in the queue as well as any new arrivals. This situation becomes more severe when both the actual job processing time and the level of underestimation increase. However, when the job processing times are generated from lower variance distributions, the underestimation of large jobs will not cause severe performance degradation \cite{mailach2017robustness}.

The Shortest Estimated Processing Time (SEPT) policy is a version of the Shortest Processing Time (SPT) policy that skips updating the estimated remaining processing times and prioritizes jobs based only on their estimated processing times. Experimental results show that SEPT has impressive performance in the presence of estimated job processing times, as well as being easier to implement than SERPT \cite{dell2019scheduling}. 

In this paper, we will discuss the problem of single-server scheduling when only estimates of the job processing times are available. In Section \ref{LiteratureReview}, we discuss the existing literature for scheduling policies that handle inexact job processing time information. Most of the existing literature analyzes and introduces size-based policies when the estimation error is relatively small, restricting applicability of the results. Furthermore, many simulation-based examinations only consider certain workload classes and are not validated over a range of job processing times and estimation error distributions. We propose a scheduling policy that exhibits desirable performance over a wide range of job processing time distributions, estimation error distributions, and workloads.

The Gittins' Index policy \cite {gittins1979bandit}, a dynamic priority-based policy, is optimal in minimizing the MST in an M/G/1 queue \cite{aalto2009gittins}. When there are job processing time estimates, the Gittins' Index policy utilizes information about job estimated processing time, and the job processing time and estimation error distributions to decide which job should be processed next. The assumption of knowing these distributions before scheduling may be problematic in real environments. Furthermore, scheduling jobs using the Gittins' Index policy introduces computational overhead that may be prohibitive. While there are significant barriers to implementing the Gittins' Index policy, our proposed policy is motivated by the form of the Gittins' Index policy.

We make the following contributions: While the SEPT policy performs well in the presence of estimated job processing times \cite {dell2019scheduling}, we first introduce a heuristic that combines the merits of SERPT and SEPT.
Secondly, we specify the Gittins' Index policy given multiplicative estimation errors and restricted to knowing only the estimation error distribution. We show that our proposed policy, which we call the Size Estimate Hedging (SEH) policy, has performance close to the Gittins' Index policy. Similar to SERPT and SEPT, the SEH policy only uses the job processing time estimates to prioritize the jobs. Finally, we provide numerical results obtained by running a wide range of simulations for both synthetic and real workloads. The key observations suggest that SEH outperforms SERPT except in scenarios where the job processing time variance is extremely low. SEH outperforms SEPT whether the variance of the job processing times is high or low. With the presence of better estimated processing times in the system (low variance in the estimation errors), SEH outperforms SEPT and has performance close to the optimal policy (SRPT) if the estimation errors are removed. On the other hand, we observe that when the estimation errors have high variance, there is little value in using the estimated processing times. We also notice that the system load does not significantly affect the relative performance of the policies under evaluation. The SEH policy treats underestimated and overestimated jobs fairly, in contrast with other policies that tend to favor only one class of jobs. When the job processing time variance is high, the SEH and SEPT policies obtain a near-optimal mean slowdown value of $1$, indicating that underestimated large jobs do not delay small jobs. In terms of mean slowdown, SEH outperforms SEPT across all levels of job processing time variance.

The rest of the paper is organized as follows. Section \ref{LiteratureReview} presents the existing literature in scheduling single-server queues with estimated job processing times. Section \ref{SEH} defines our SEH policy and discusses its relationship to a Gittins' Index approach. Our simulation experiments are described in detail in Section \ref {EvaluationMethodology}.
We provide the results of our simulations in Section \ref{SimulationResultsplusInsights} and conclude and discuss future directions in Section \ref{Conclusion}.
\section{Related Work} \label{LiteratureReview}

Scheduling policies and their performance evaluation in a preemptive M/G/1 queue have been a subject of interest for some time. Size-based policies are known to perform better than size-oblivious policies with respect to sojourn times. In fact, the SRPT policy is optimal in minimizing the MST \cite{schrage1968letter}. However, size-based policies have a considerable disadvantage: When the exact processing times are not known to the system before scheduling, which is often the case in practical settings, their performance may significantly degrade. Dell’Amico et al.\ \cite{dell2015psbs} study the performance of SRPT with estimated job processing times and demonstrate the consequences of job processing time underestimations under different settings. Studies in Harchol-Balter et al.\ \cite{harchol2003size} and Chang et al.\ \cite{chang2011scheduling} discuss the effect of inexact processing time information in size-based policies for web servers and MapReduce systems, respectively. Our paper assumes that the processing time is not available to the scheduler until the job is fully processed, but that processing time estimations are available. The related literature for this setting is reviewed in the following paragraph.

Lu et al.\ \cite{lu2004size} were the first to study this setting. They show that size-based policies only benefit the performance when the correlation between a job’s real and estimated processing time is high. The results in Wierman and Nuyens \cite{wierman2008scheduling}, Bender et al.\ \cite{bender2002improved}, and Becchetti et al.\ \cite{becchetti2004semi} are obtained by making assumptions that may be problematic in practice. A strict upper bound on the estimation error is assumed in \cite{wierman2008scheduling}. On the other hand, \cite{becchetti2004semi} and \cite{bender2002improved} define specific job processing time classes and schedule the jobs based on their processing time class, which can be problematic for very small or very large jobs. This setting is also known as semi-clairvoyant scheduling. In this work, we do not assume any bounds on the estimation error or assign jobs to particular processing time classes. Consistent with this body of work, we do find that SEH is not recommended for systems with large estimation error variance. However, we do find that it performs well for levels of estimation error variance that are typically found in practice.

When the job processing time distribution is available, the Gittins’ Index policy \cite{gittins1979bandit} assigns a score to each job based on the processing time it has received so far, and the scheduler chooses the job with the highest score to process at each point in time. This policy is proven to be optimal for minimizing the MST in a single-server queue when the job processing time distribution is known \cite{aalto2009gittins}. This policy is specified in the next section.

\section{Size Estimate Hedging: A Simple Dynamic Priority Scheduling Policy} \label{SEH}

\subsection{Model} \label{GeneralModel}
Consider an M/G/1 queue where preemption is allowed and we are interested in minimizing the MST. 
We assume that a job's processing time is not known upon arrival; however, an estimated processing time is provided to the scheduler. We concentrate on a multiplicative error model where the error distribution is independent of the job processing time distribution. The estimated processing time 
$\hat S$ of a job is defined as $\hat S = SX$ where $S$ is the job processing time and $X$ is the job processing time
estimation error. We assume that the value of $\hat S$ is known upon each job's arrival and is denoted by $\hat{s}$. 
The choice of a multiplicative error model results in having an absolute error proportional to the job processing time $S$, thus avoiding situations where the estimation errors tend to be worse for small jobs than for large jobs. Furthermore, Dell'Amico et al.\ \cite{dell2015psbs} and Pastorelli et al.\ \cite{pastorelli2013hfsp} suggest that a multiplicative error model is a better reflection of reality. To define our scheduling policies, we also require the notion of a quantum of service. The job with the highest priority is processed for a quantum of service $\Delta$ until either it completes or a new job arrives. At that point, priorities are recomputed.

\subsection{Gittins' Index Approach} \label{GittinsIndex}
The Gittins' Index Policy is an appropriate technique for determining scheduling policies when the job processing time and estimation error distributions are known. For a waiting job $i$, an index $G(a_i)$ is calculated, where $a_i$ is the elapsed processing time. At each time epoch, the Gittins' Index policy processes the job with the
highest index $G(a)$ among all of the present waiting jobs \cite{gittins1979bandit}. The Gittins' rule takes the job's elapsed processing time into account and calculates the optimal quantum of service $\Delta ^*(a)$ that it should receive.

The associated efficiency function $J(a,\Delta ),{\rm{ }}a,{\rm{ }}\Delta  \ge 0$ of a job
with processing time $S$, elapsed processing time $a$ and quantum of service $\Delta $ is defined as

\begin{equation}\label{eq:1} J(a,\Delta ) = \frac{{P(S - a \le \Delta |S > a)}}{{E[\min \{ S - a,{\rm{
}}\Delta \} |S > a]}}.
\end{equation}

The numerator is the probability that the job will be completed within a quantum of service $\Delta $, and the denominator is
the expected remaining processing time a job with elapsed processing time $a$ and quantum of service
$\Delta $ will require to be completed.

The server (preemptively) processes the job with the highest
index at each decision epoch. Decisions are made when  \begin {enumerate*}  \item a
new job arrives to the queue, \item the current job under processing completes, or \item
the current job receives its optimal quantum of service and
does not complete.\end {enumerate*} If there are multiple jobs that have the same highest
index and all have zero optimal quanta of service, the processor will be shared
among them as long as this situation does not change. If there is only one job
with the highest index and zero optimal quantum of service, its index should be
updated throughout its processing \cite {aalto2009gittins}. 

Although the Gittins' Index policy is optimal in terms of minimizing the mean
sojourn time in an $M/G/1$ queue \cite {aalto2009gittins}, the assumption of knowing the job size and estimation error distributions might not always be practical to make. Furthermore, forming the Gittins' Index policy's efficiency function has significant computational overhead. As a result, this policy may be a problematic choice for real environments where the scheduling speed is important. However, examining the form of optimal policies has helped us in the construction of a simple heuristic. In particular, the notion of defining a policy in terms of an index allows us to make precise our notion of combining the relative merits of SRPT and SEPT. 

\subsection{Motivation} \label{Motivation}
When a job enters the system under SERPT, there is no basis on which to assume that the estimated processing time, $\hat{s}$, is incorrect. However, when the elapsed processing time reaches $\hat{s}$, we are certain that the job processing time has been underestimated. In addition, Dell'Amico et al.\ \cite{dell2019scheduling} show that SEPT performs well when dealing with estimated processing times and in the presence of estimation errors, in particular severe underestimates. So, we would like to combine these two policies. A convenient way to do this is to introduce a Gittins'-like score function, where a higher score indicates a higher priority. We will be aggressive and use the score function for SERPT until the point that we know a job is underestimated and then freeze the score, which is similar to what SEPT's constant score function does (see \eqref{eq:11} below). In this way, instead of switching to SEPT's score function, we would like to give credit for the jobs' cumulative elapsed processing times.

The score functions for SRPT, SERPT, and
SEPT are provided in \eqref{eq:9}, \eqref{eq:10}, and \eqref{eq:11}, respectively.

\begin {equation}\label{eq:9}
G(a,s)=\frac{1}{{s - a}},
\end{equation}

\begin {equation}\label{eq:10}
G(a,\hat{s})=
\begin{cases}
\frac{1}{{\hat s - a}}, &  \hat s > a ,\\
\infty,  &  \hat s \le a,
\end{cases}
\end{equation}

\begin {equation}\label{eq:11}
G(a,\hat{s})=\frac{1}{{\hat s}}.
\end{equation}

We note that \eqref{eq:9} and \eqref{eq:10} have an increasing score function, and \eqref{eq:11} always assigns a constant score for a particular job.

\subsection{The SEH Policy} \label{FormalDefiniton}

Combining the score functions for SERPT and SEPT, we now define our policy. As discussed in the previous section, we would like to transition between SERPT when we cannot determine if a job processing time is underestimated to a fixed priority like SEPT when it is determined that underestimation has occurred. One consequence of using this policy is that any underestimated small job can still receive a ``high'' score and be processed, while underestimated large jobs will have a much lower score and do not interfere, even with underestimated small jobs.  Furthermore, not needing to know the job processing time and estimation error distribution, the SEH Policy does not have much overhead. Thus, it can schedule the jobs at a speed comparable to the SEPT policy. 

We introduce the score function of our SEH policy as

\begin {equation}\label{eq:8}
G(a,\hat{s})=
\begin{cases}
\frac{1}{{\hat s - a(1 - \frac{a}{{2\hat s}})}}, & 0 \le a < \hat s,\\
\frac{2}{{\hat s}}, & a \ge \hat s,
\end{cases}
\end{equation}

\noindent where the scheduling decisions are only made at arrivals and departures. 

With the score function in \eqref{eq:8}, a job's score will increase up to the point that
it receives processing equal to its estimated processing time and then receives a constant score of
$\frac{2}{{\hat s}}$ until it completes. The choice of $2$ was made after some experimentation, it would be worthwhile to explore the sensitivity of the performance to this choice.

\subsection{Gittins' Index vs. SEH} \label{GittinsVsSEH}
In this section, we show that the form of our policy is consistent with the Gittins' index in the setting that we only know the error estimate distribution. In particular, we have no a priori or learned knowledge of the processing time distribution.

With our estimation model in mind, \eqref{eq:1} can be rewritten as 

\begin{equation} \label{eq:2} J(a,\Delta,\hat{s} ) = \frac{{P(\frac{{\hat s}}{X} - a \le \Delta |\frac{{\hat s}}{X} >
a)}}{{E[\min \{ \frac{{\hat s}}{X} - a,{\rm{ }}\Delta \} |\frac{{\hat s}}{X} >
a]}}.
\end{equation}
The Gittins' index $G(a,\hat{s}),{\rm{ }}a \ge 0,$ is defined by 
\begin{equation*}\label{eq:3}
G(a,\hat{s}) = \mathop {\sup
}\limits_{\Delta  \ge 0} {\rm{ }}J(a,\Delta,\hat{s} ).
\end{equation*}
The optimal quantum of service is
denoted as 
\begin{equation*}\label{eq:4}
{\Delta ^*}(a,\hat{s}) = \sup \{ \Delta  \ge 0|G(a,\hat{s}) = J(a,\Delta ,\hat{s} )\}.
\end{equation*}

Suppose that the lower and upper limits on the estimation error distribution are $l$ and $u$, respectively ($l$ may be zero and $u$ may be $\infty$). After some calculation, the Gittins' index can then be written as
\begin{equation}\label{eq:6}
G(a,\hat{s}) = 
\begin{cases}

\frac{1}{{\hat s - aE[X|X \le \frac{{\hat s}}{a}]}}, & \frac{{\hat s}}{a} < u,\\ 
\frac{1}{{\hat s - aE[X]}},& otherwise,
\end{cases}
\end{equation}
\noindent where ${\Delta ^*}  = \frac{{\hat s}}{l} - a$. For instance, the Gittins' index for a $Log - N(\mu,{\rm{
}}{\sigma ^2})$ error distribution is
\begin {equation}\label{eq:7}
G(a,\hat{s}) = \frac{1}{{\hat s - a{e^{\mu  + g(a,\hat{s})}}}}, \end{equation}

\noindent where 
\begin {equation*}\label{eq:12}
g(a,\hat{s})= \frac{{{\sigma ^2}\phi[ \frac{{\ln
(\frac{{\hat s}}{a}) - \mu  - {\sigma ^2}}}{\sigma }]}}{{{2}\phi [\frac{{\ln
(\frac{{\hat s}}{a}) - \mu }}{\sigma }]}},
\end{equation*}
\noindent and $\phi $ is the cumulative distribution function of the $Log - N(0,{\rm{
}}{\sigma ^2})$ distribution. Note that for the Log-Normal distribution as the job processing time error distribution, the second case in \eqref{eq:6} cannot happen. For the remainder of the paper, we will refer to this policy as the Gittins' Index policy. We recognize that this is a slight abuse of terminology, as we are ignoring the job processing time distribution.

\begin{figure}[t]
\centering

\subfigure[calculated with the Gittins' Index policy]{
\includegraphics[width=0.47\textwidth]{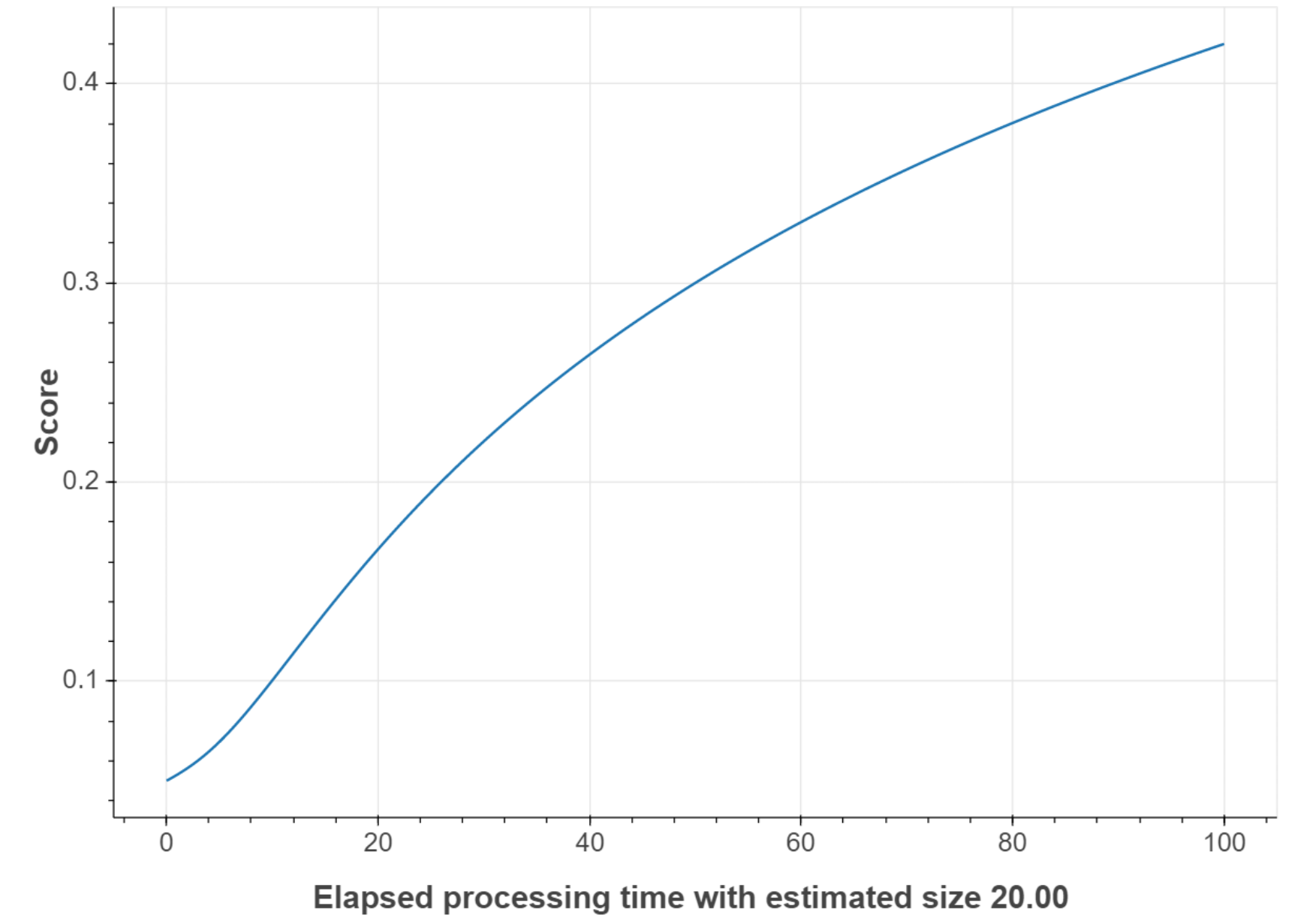}\label{fig:Gittins}
}
\subfigure[calculated with SEH]{
\includegraphics[width=0.47\textwidth]{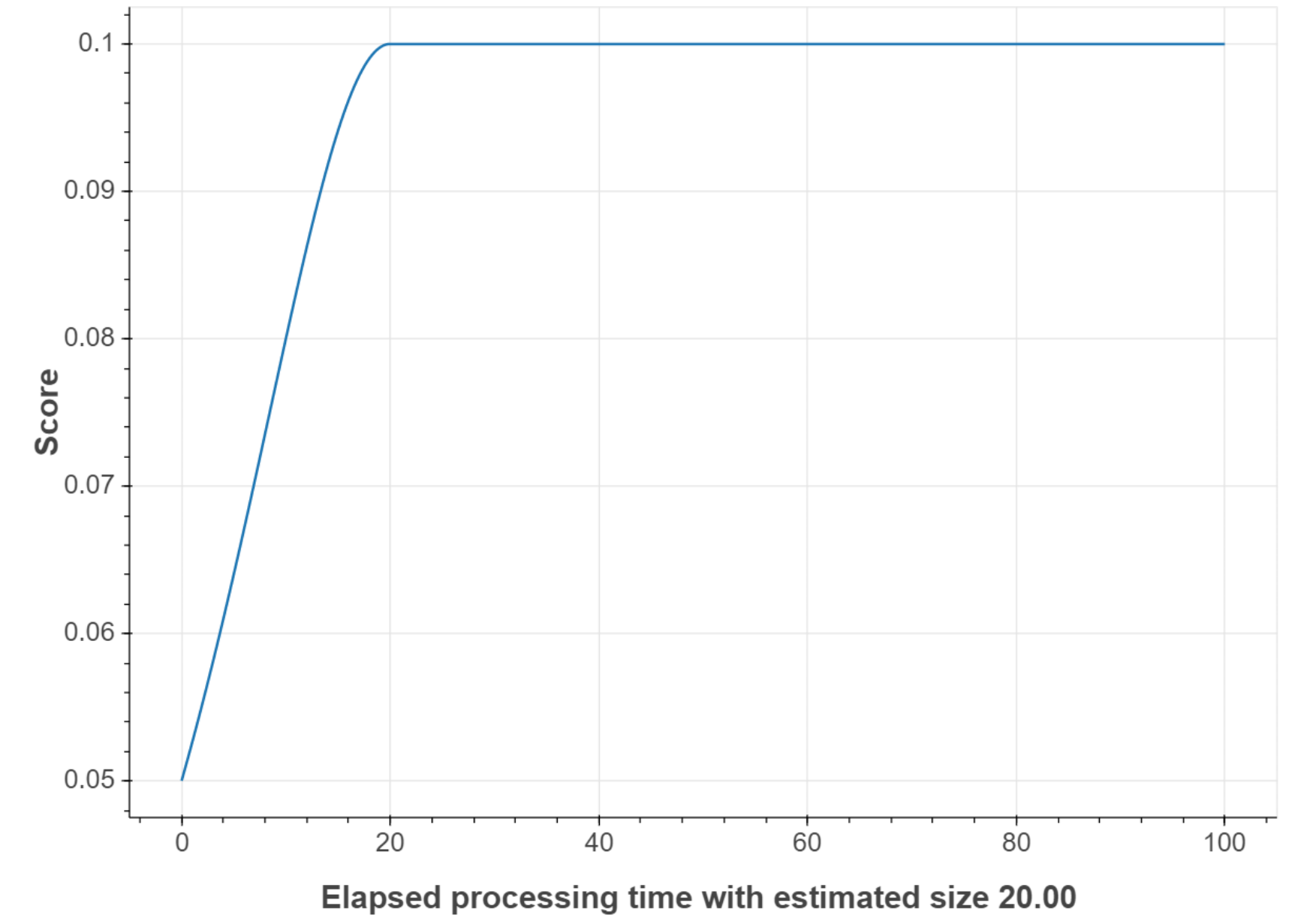} \label{fig:NP}
}
\caption{\label{fig:Gittins_VS_SEH} Job score as a function of the elapsed processing time } 
\end{figure}

Taking the score in \eqref{eq:7} into account, for any job with an estimated processing time $\hat s$, the score calculated with the Gittins' Index policy continuously increases until the job completes. Fig. \ref{fig:Gittins} shows this score for a job with an estimated processing time of 20 and an estimation error generated from a $Log - N(0,{\rm{}}{\sigma ^2})$ distribution as a function of its elapsed processing time. We observe that for larger values of elapsed processing time, the slope of the score is decreasing. Fig. \ref{fig:NP} shows the score calculated with the SEH policy for a job with an estimated processing time of 20 as a function of its elapsed processing time. The score shown in Fig. \ref{fig:Gittins} is consistent with the score function having decreasing slope at some point beyond the point at which the elapsed processing time reaches the estimated processing time, as in Fig. \ref{fig:NP}. Of course, the change in slope for SEH is more severe, but we will see in our numerical experiments that the performance of the two policies is quite close. SEH has less computational overhead and more importantly, does not require knowledge of the estimation error distribution.

\section{Evaluation Methodology} \label{EvaluationMethodology}
\subsection{Policies Under Evaluation} \label{PoliciesUnderEvaluation}

In this section, we introduce the size-based scheduling policies considered for
evaluation. As our baseline policy, we consider the SRPT policy when the exact
job processing times, given by $s$, are known before scheduling. The SRPT policy is an ``ideal'' policy since it assumes that there are no errors in estimating the processing time. 
\begin{itemize}

\item \textbf{SERPT policy} --- The SERPT policy is a version of SRPT that uses the estimates of job processing times as if they were the true processing times.

\item \textbf{SEPT policy} --- The SPT policy skips the SRPT policy's updating of remaining processing times and only schedules jobs based on their estimated processing time. 

\item \textbf{SEH and Gittins' Index policy} --- Our proposed SEH policy and the Gittins' Index policy are explained in detail in Section \ref{FormalDefiniton} and Section \ref{GittinsVsSEH}, respectively. 
\end{itemize}

All these policies fit into the ``scoring'' framework, and they assign scores to each job
and process the jobs in the queue in the descending order of their scores.
Moreover, pre-emption is allowed, and a newly-arrived job can pre-empt the
current job if it has a higher score. The score functions in  \eqref{eq:9},  \eqref{eq:10}, \eqref{eq:11}, \eqref{eq:8}, and \eqref{eq:6} show how we
calculate the scores for the SRPT, SERPT, SEPT, SEH, and Gittins' Index policy,
respectively. 

\subsection{Performance Metrics} \label{PerformanceMetrics}

We evaluate the policies defined in Section \ref{PoliciesUnderEvaluation} with respect to two performance metrics: MST and Mean Slowdown. When the job processing times have large variance, the sojourn times for small jobs and large jobs differ significantly. Thus, we use the per job slowdown, the ratio between a job's sojourn time and its processing time \cite{wierman2011fairness}.

\subsection{Simulation Parameters} \label{SimulationParameters}
We would like to evaluate the policies over a wide range of job processing time and error distributions. To generate this range of distributions, we fix the form of the distribution and vary the parameters. We use the same settings that Dell'Amico et al.\ \cite {dell2015psbs} use in their work. Table \ref{tab:1} provides the default parameter values that we use in our simulation study. We now provide details of our simulation model. Note that our policy fits into the SOAP framework of Scully et al.\ \cite{scully2018soap}, however as we are also evaluating mean slowdown, we chose simulation for evaluation.\\

\begin{table}[t]
\centering
\caption{Parameter Settings}
\begin{tabular}{ |p{2.2cm}|p{7cm}|p{2cm}|  }
 \hline

 \hline
 \textbf{Parameter} & \textbf{Definition} &\textbf{Default} \\
 \hline
 \# jobs   & the number of departed jobs     &$10,000$ \\ 
 $k$&   shape for Weibull job processing time distribution  & $0.25$   \\
 $\sigma$ & $\sigma$ in the Log-Normal error distribution & $0.5$ \\
 $\rho$ &   system load   & $0.9$ \\

 \hline

\end{tabular}

\label{tab:1}
\end{table}

\textbf{Job Processing Time Distribution} ---  We consider an $M/G/1$ queue where the processing time is generated according to a Weibull distribution. This allows us to model high variance processing time distributions, which better reflect the reality of computer systems (see \cite{crovella1998heavy} and \cite{harchol1999ect} for example).  In general, the choice of a Weibull distribution gives us the flexibility to model a range of scenarios. The shape parameter $k$ in the Weibull distribution allows us to evaluate both high variance (smaller $k$) and low variance (larger $k$) processing time distributions.  

Considering that the job processing time distribution plays a significant role in the scheduling policies' performance and size-based policies show different behaviors with high variance job processing time distributions, we choose $k=0.25$ as our default shape for the Weibull job processing time distribution. With this choice for $k$, the scheduling policies' performance is highly influenced by a few very large jobs that constitute a substantial percentage of the system's overall workload. 
We vary $k$ between $0.25$ and $2$, considering specific values of $0.25$, $0.375$, $0.5$, $0.75$, $1$, and $2$. We show that the SEH policy performs best in the presence of high variance job processing time distributions. \\

\textbf{Job Processing Time Error Distribution} --- We have chosen the Log-Normal distribution as our error distribution so that a job has an equal probability of being overestimated or underestimated. The Gittins' index for this estimation error distribution is shown in \eqref{eq:7}. The $\sigma$ parameter controls the correlation between the actual and estimated processing time, as well as the estimation error variance. By increasing the $\sigma$ value,  the correlation coefficient becomes smaller, and the estimation error variance increases, resulting in the occurrence of more large underestimations/overestimations (more imprecise processing times). We choose $\sigma=0.5$ as the default value that corresponds to a median relative error factor of $1.40$. We vary $\sigma$ between $0.25$ and $1$ with specific values of $0.25$, $0.375$, $0.5$, $0.75$, and $1$ to better illustrate the effect of $\sigma$ on the evaluated performance. \\

\textbf{System Load} --- Following Lu et al.\ \cite{lu2004size}, we consider $\rho = 0.9$ as the default load value and vary $\rho$ between $0.5$ (lightly loaded) and $0.95$ (heavily loaded) with increments of 0.05 and an additional system load of $0.99$. \\

\textbf{Number of Jobs} --- The number of jobs in each simulation run is $10,000$ and a simulation run ends when the first $10,000$ jobs that arrived to the system are completed. We fix the confidence level at $95\%$, and for each simulation setting, we continue to perform simulation runs until the width of the confidence interval is within $5\%$ of the estimated value. For low variance processing time distributions (larger $k$), $30$ simulation runs suffice; however, more simulation runs are required for high variance processing time distributions (smaller $k$).

\section{Simulation Results} \label{SimulationResultsplusInsights}
In this section, we evaluate the performance of the policies in Section \ref{PoliciesUnderEvaluation} by running experiments on both synthetic and real workloads. We run different simulations by generating synthetic workloads based on different job processing time and error parameters and we analyze these parameters' effect on the performance of each of the policies.

For evaluating our results in practical environments, we consider a real trace from a Facebook Hadoop cluster in 2010 \cite{chen2012interactive} and show that the policies' performance is consistent with the results we obtained with synthetic workloads.
The key observations, validated both on synthetic and real workloads, are highlighted as follows:

\begin{itemize}
\item	The Gittins’ Index policy outperforms SERPT for all the evaluated values of $k$ and $\sigma$. We show the same observation with our proposed SEH policy except for values of $k$ that correspond to very low job processing time variance.
\item The Gittins’ Index and SEH policies outperform SEPT with lower values of $\sigma$ (better estimated processing times) and have an MST near the optimal MST obtained without any estimation errors. 
\item SEH performs well in reducing both the MST of overestimated jobs and underestimated jobs. 
\item	The load parameter does not have a significant effect on the relative values of the MST obtained with the evaluated policies.
\item	The Gittins' Index, SEH and SEPT policies have a near-optimal mean slowdown of 1 when the estimated processing
times have high variance.
\item	The SEH performs best across all values of $k$ in terms of minimizing the mean slowdown.

\end{itemize} 

In what follows, we discuss the numerical results and how they support these key observations. \\

\textbf{Synthetic Workloads} --- We first note that the job processing time $k$ parameter and the estimation error $\sigma$ parameter have the greatest impact on the policies' performance. Thus, we focus on varying these parameters. We show that the Gittins' Index policy outperforms SERPT across all evaluated values of $k$ and $\sigma$ and our SEH policy outperforms SERPT except for the values of $k$ and $\sigma$ that correspond to distributions with extremely low variance. For the scenarios where we do not state the parameter values explicitly, the parameters in Table \ref{tab:1} (see Section \ref{SimulationParameters}) are considered. 

\begin{figure}[ht]

  \includegraphics[width=100mm,scale=0.5]{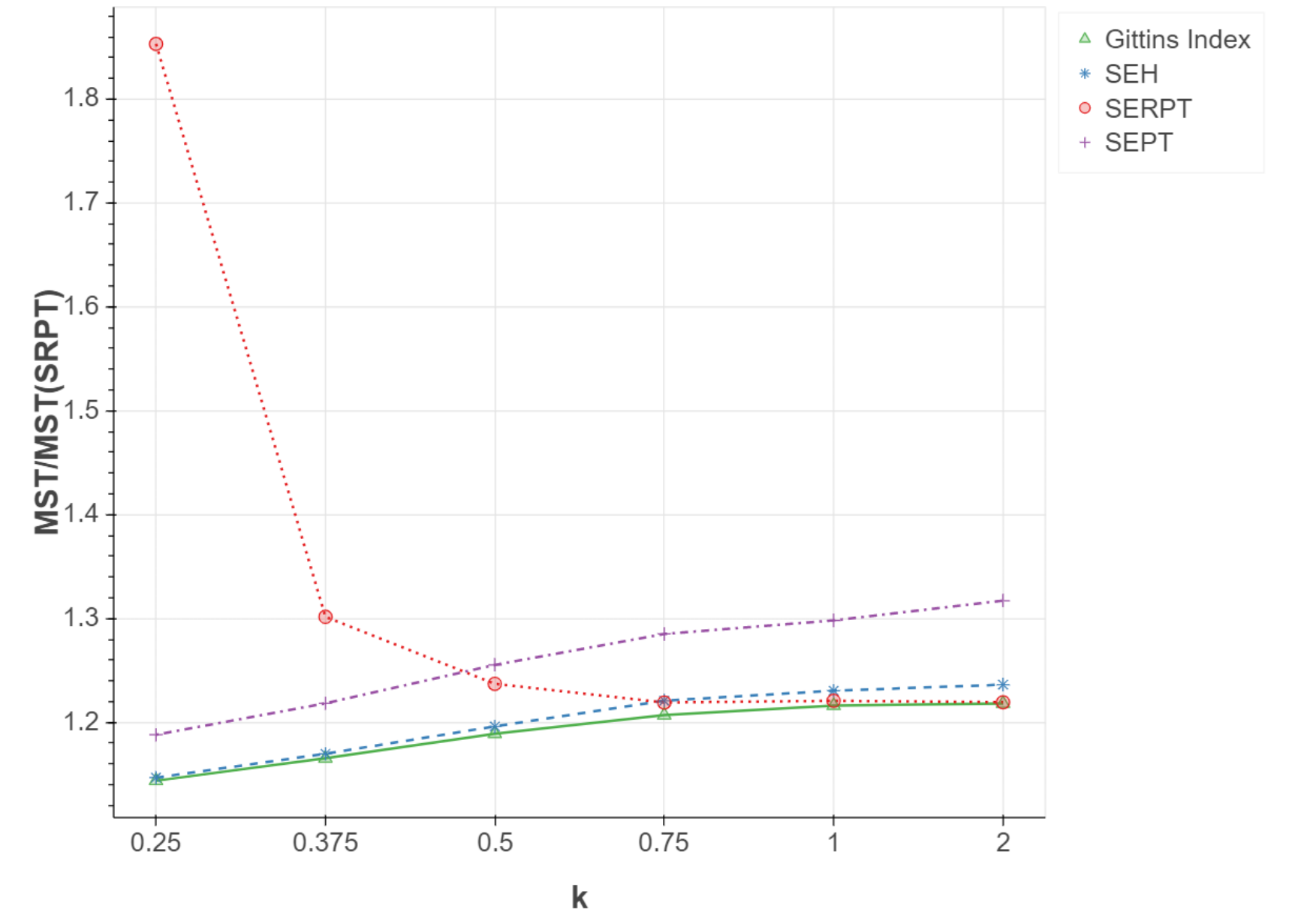}
  \centering
    \caption{Impact of $k$ on the MST}
  \label{fig:5ShapeMST}

\end{figure}

Fig. \ref{fig:5ShapeMST} captures the impact of job processing time variance and displays the MST of the Gittins' Index, SEH, SERPT, and SEPT policies normalized against the MST obtained with SRPT with $\sigma$ having the default value of $0.5$. We observe that for a high variance job processing time distribution ($k = 0.25$), SERPT performs very poorly compared to the other policies due to the presence of large, underestimated jobs. We note that the SERPT policy performs well if the variance of the processing times is sufficiently low. Based on Fig. \ref{fig:5ShapeMST}, we notice that the gap between SEPT and the Gittins' Index policy grows slightly when the job processing time variance is lower. The gap between SEH and the Gittins’ Index policy also grows but not to the same degree as SEPT. For $k > 0.75$, the performance of the Gittins' Index policy, SEH, and SERPT are quite close. In fact, we observe that our SEH policy performs very close to the Gittins' Index policy across all values of $k$. Furthermore, we notice that as the variance in processing times gets smaller, the gap between what is achievable by the policy under evaluation and what is achievable if there were no errors is larger than for the high variance scenarios. 

\begin{figure}[ht]

  \includegraphics[width=100mm,scale=0.5]{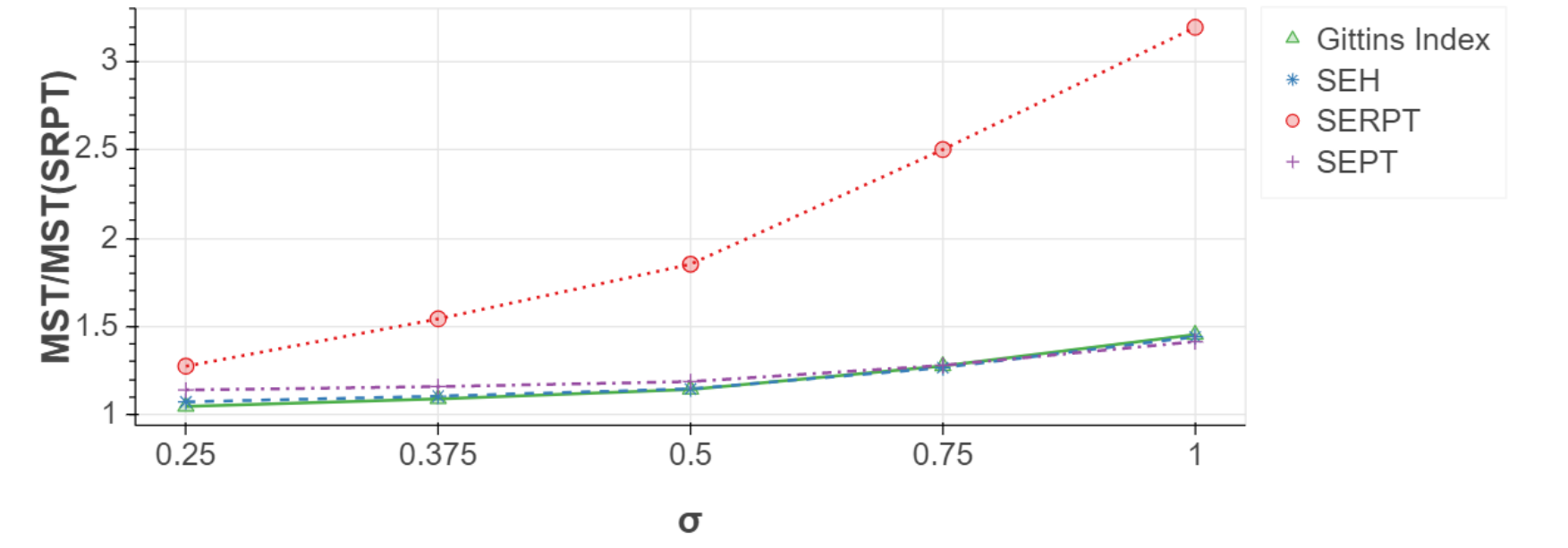}
  \centering
    \caption{Impact of $\sigma$ on the MST}
  \label{fig:25SigmaMST}

\end{figure}

The shape parameter $k$ affects the job processing time variance and the scheduling policies' performance the most, especially when the job processing time distribution has high variance. We can be optimistic about using estimates if the variance is low, but we have to be careful in choosing the scheduling policy if the job processing time variance is high. The literature focuses on high variance workloads, and we will continue evaluating the policies on such workloads. In Fig. \ref{fig:25SigmaMST}, we display the normalized MST of the policies against the MST of the SRPT policy under varying $\sigma$, $ \rho = 0.9$, and the default $ k = 0.25$. We notice that the Gittins' Index, SEH, and SEPT policies are relatively insensitive to the $\sigma$ value, while the gap between these three policies and SERPT increases with increasing $\sigma$. In fact, the Gittins' Index and SEH policies outperform SEPT with $\sigma \le 0.5$ and have an MST near the optimal MST obtained without any estimation errors. We conclude that the impact of the Gittins' Index policy and SEH becomes more prominent when the estimates improve.

In Fig. \ref{fig:25SigmaMST}, we observe that while choosing a more aggressive policy like the Gittins' Index and SEH policies is a good choice under lower values of $\sigma$, SEPT is preferred when $\sigma = 1$. The reason is that lower values of $k$ (here, $k=0.25$), cause more large jobs in the system. Furthermore, for values of ${\sigma \ge 1 }$, the estimation errors have high variance and thus the estimated processing
times can be very imprecise. We notice that both SEH and the Gittins' Index policy suffer from a slight promotion of severely underestimated jobs that leads to temporary blockage for the other jobs. What has happened in this case is that the estimates of the processing times have degraded to the point that they are not useful. In particular, one should instead base scheduling decisions on the processing time distribution, so for example in scenarios with high variance in both processing times and estimation errors, a policy which ignores the estimates, such as Least Attained Service (LAS) would be warranted. The LAS scheduling policy \cite {rai2003analysis}, also known as Shortest Elapsed Time \cite {coffman1973operating} and Foreground-Background \cite{kleinrock1975theory}, pre-emptively prioritizes the job(s) that have been processed the least. If more than one job has received the least amount of processing time, the jobs will share the processor in a processor-sharing mode. Analytic results in \cite{righter1989scheduling} and \cite {yashkov1987processor} show that LAS minimizes MST when the job processing time distribution has a decreasing hazard rate and there are no processing time estimates available.

These observations are consistent with the results in Table \ref{tab:2} which considers the same settings as in Fig. \ref{fig:25SigmaMST} when $\sigma = 1$ and $\sigma = 2$. SERPT has poor performance compared the other policies under $\sigma \ge 1$ and thus is not included. Pastorelli et al.\ \cite{pastorelli2013hfsp} show that lower values of $\sigma$ ($\sigma < 1$) are what one sees in practice. It would be interesting to look at the optimal Gittins' Index policy that includes both the job processing time and estimation error distributions, as it would capture this effect. Although doing so can help develop policies that are effective even at high values of $\sigma$, deriving the Gittins' index would be quite complicated with this extra condition, but it could give insight into designing simpler policies.

\begin{table}[t]
\centering
\caption{Policies evaluation under $\sigma=1$ and $\sigma=2$}
\begin{tabular}{|p{2cm}|p{2.30cm}|p{2.30cm}|p{2.30cm}|p{2.30cm}|} 
\hline 
\multirow{2}{*}{} & \multicolumn{2}{|p{1.2cm}|}{$\bm{\sigma=1}$}  & \multicolumn{2}{|p{1.2cm}|}{$\bm{\sigma=2}$}  \\ \cline{2-5} 
\textbf{Policy} & MST/ MST(SRPT) & Mean Slowdown & MST/ MST(SRPT) & Mean Slowdown \\ \hline 
Gittins' Index & 1.45 & 1.26 & 2.68 & 6.78 \\ 
SEH & 1.44 & 1.22 & 2.71 & 6.87 \\ 
SEPT & 1.41 & 1.16 & 2.54 & 4.71 \\  
LAS & 1.81 & 1.27 & 1.81 & 1.27 \\  
SRPT & 1 & 1.06 & 1 & 1.06 \\ 
\hline 
\end{tabular}

\label{tab:2}
\end{table}

\begin{figure}[t]
\centering
\subfigure[All jobs]{
\includegraphics[width=0.90\textwidth]{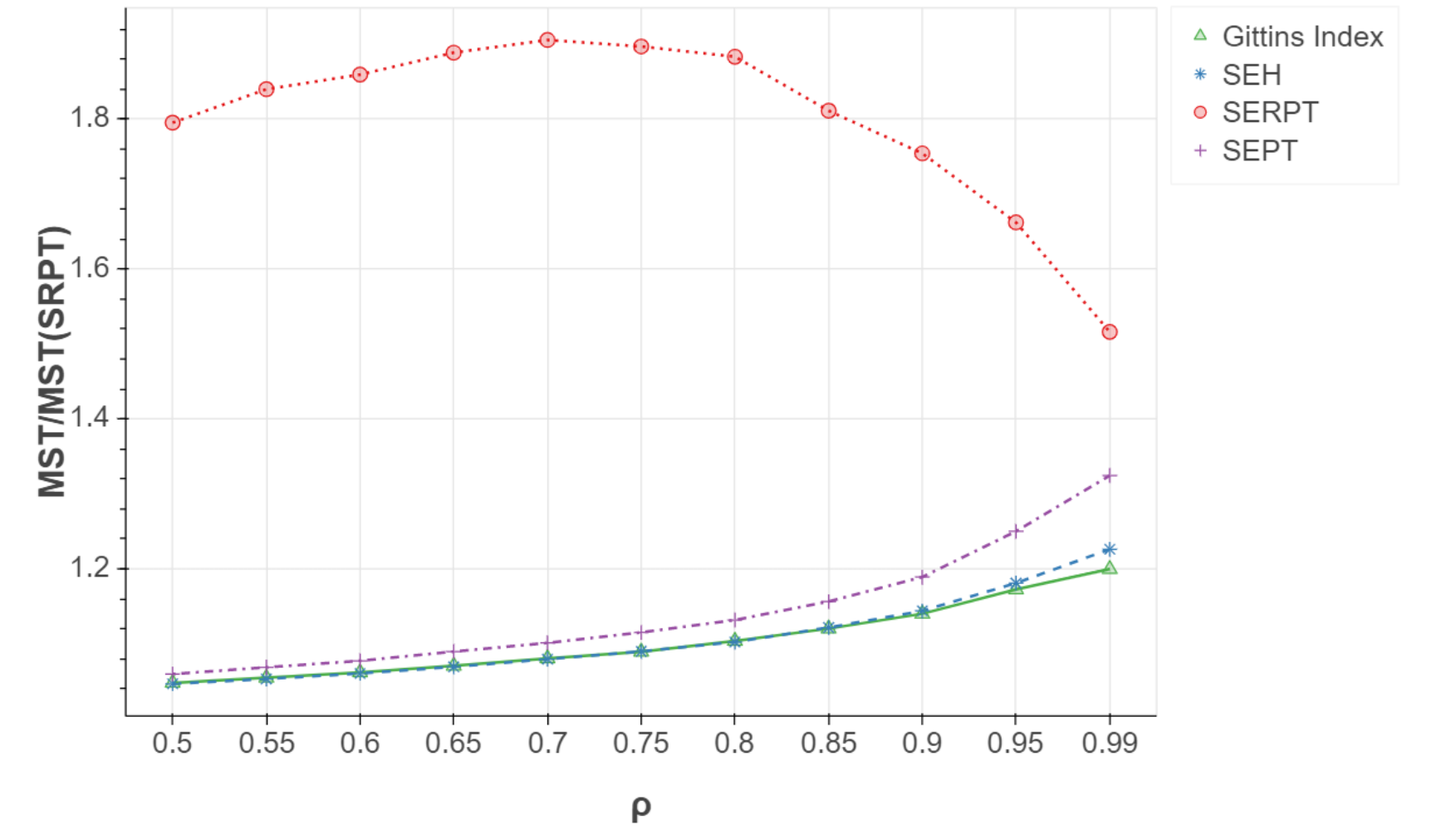}\label{fig:25Shape5SigmaMST}
}
\subfigure[Overestimated jobs]{
\includegraphics[width=0.47\textwidth]{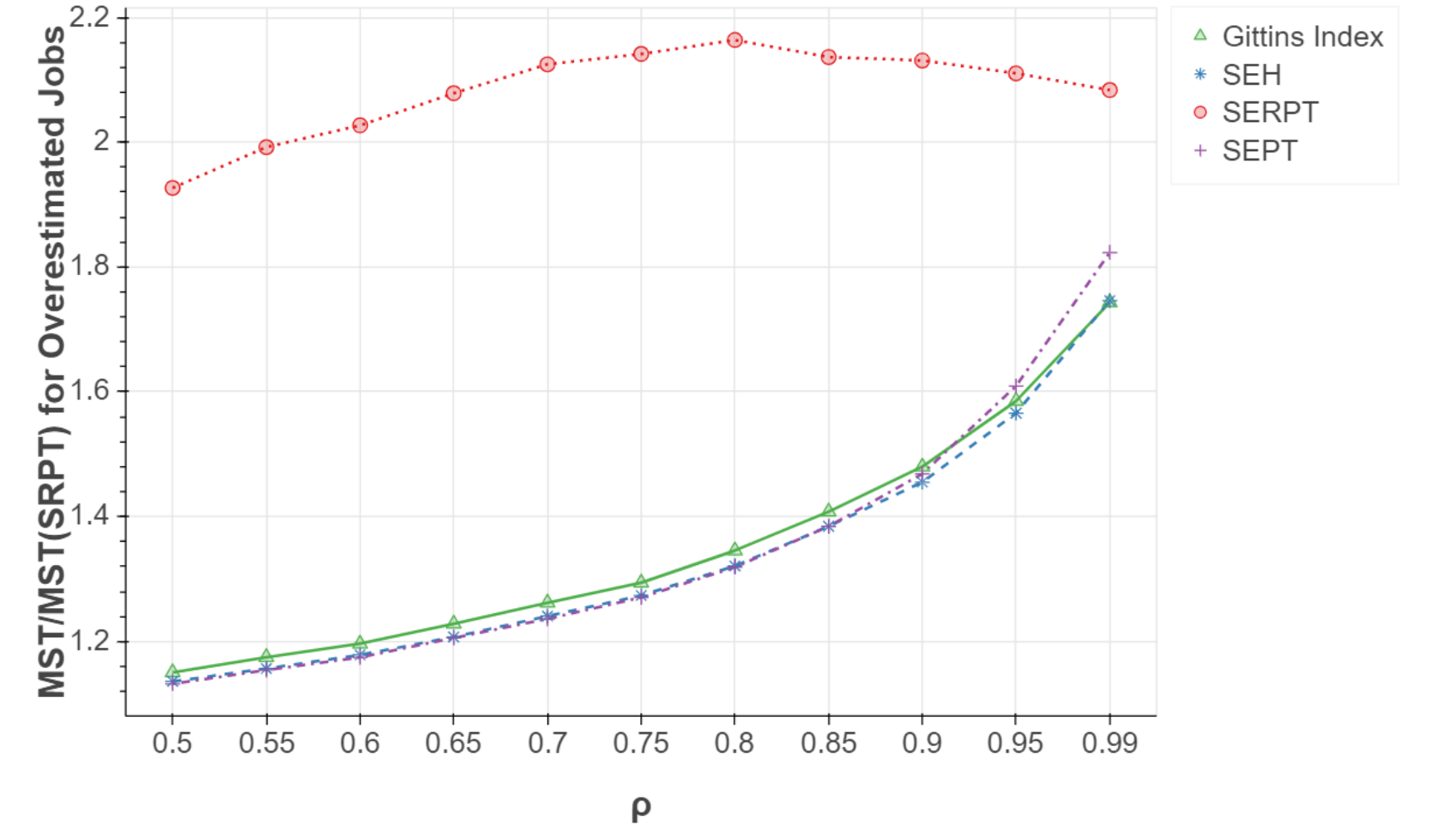}\label{fig:25Shape5SigmaMSTOV}
}
\subfigure[Underestimated jobs]{
\includegraphics[width=0.47\textwidth]{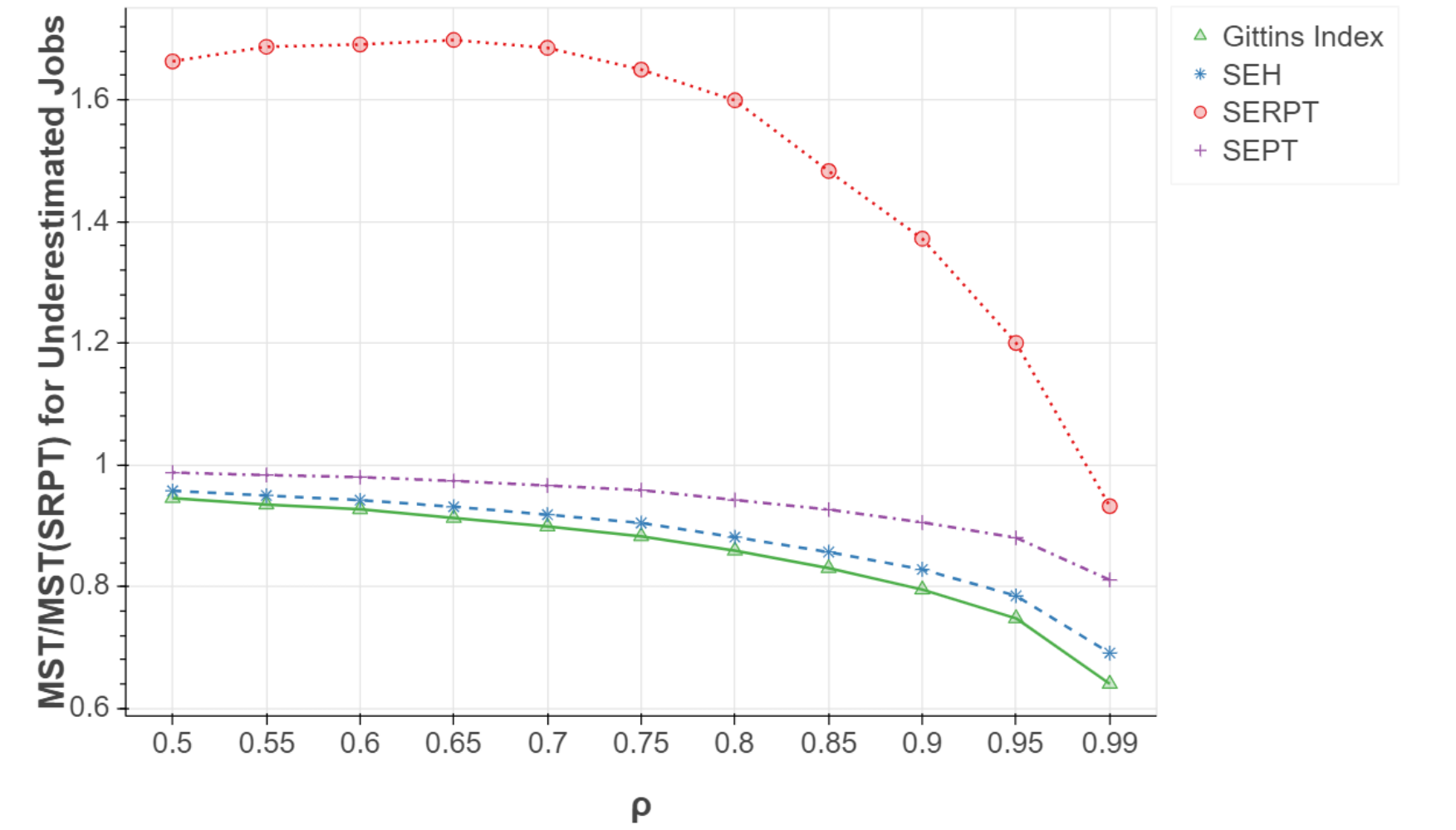} \label{fig:25Shape5SigmaMSTUN}
}
\caption{\label{fig:whole} Impact of $\rho$  on the MST} 
\end{figure}

Fig. \ref{fig:25Shape5SigmaMST}, Fig. \ref{fig:25Shape5SigmaMSTOV}, and Fig. \ref{fig:25Shape5SigmaMSTUN} show the result of simulations with the default values in Table \ref{tab:1} and varying the system load between $0.5$ and $0.99$ for all jobs, only the overestimated jobs, and only the underestimated jobs, respectively. If we concentrate only on one class of jobs (overestimated or underestimated), the policy that minimizes the MST the most can be different. We observe that the Gittins' Index and SEH policies perform best in minimizing the overall MST given different system loads. The Gittins' Index policy performs best in reducing the MST of underestimated jobs and the SEH policy has desirable performance in reducing the MST of all jobs, the overestimated jobs, and the underestimated jobs. Fig. \ref{fig:25Shape5SigmaMST} shows that the load parameter does not have a significant effect on the MST since the ratio between the MST of each policy and the MST of SRPT remains almost unchanged.

\begin{figure}[ht]

  \includegraphics[width=100mm,scale=0.5]{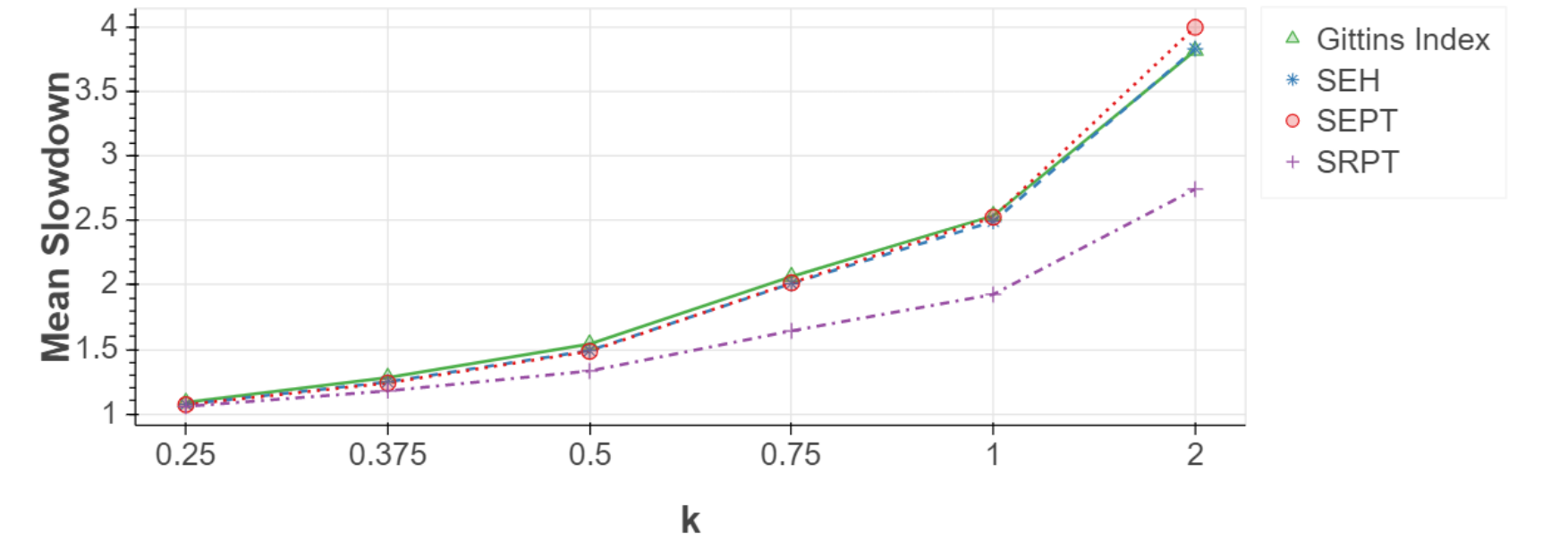}
  \centering
    \caption{Impact of $k$ on the mean slowdown}
  \label{fig:25SigmaMSD}

\end{figure}

The mean slowdown is the other metric we consider to evaluate the performance of the policies. High values of mean slowdown indicate that some jobs spend a disproportionate amount of time waiting. In Fig. \ref{fig:25SigmaMSD}, we show the mean slowdown for different values of $k$ with $\rho = 0.9$ and a $\sigma$ value of $0.5$. The mean slowdown of SERPT is not included since it is several orders of magnitude higher for $k \le 0.5$. We see that the Gittins' Index, SEH, and SEPT policies have similar performance. All policies have a near-optimal mean slowdown of 1 for high variance job processing time distributions (smaller $k$). The reason is that the very small jobs (that make up the majority of the jobs) are processed the moment they enter the system, and no large job blocks them. We also observe that SEH performs best across all values of $k$ in terms of minimizing the mean slowdown.

We conclude our experiments with synthetic workloads by indicating that the Gittins' Index and SEH policies perform better than SERPT under different parameter settings. The only exception is extreme situations like the low variance job processing time distributions (larger $k$) where SERPT outperforms SEH and works analogously to the Gittins' Index policy. \\

\begin{figure}[ht]

  \includegraphics[width=100mm,scale=0.5]{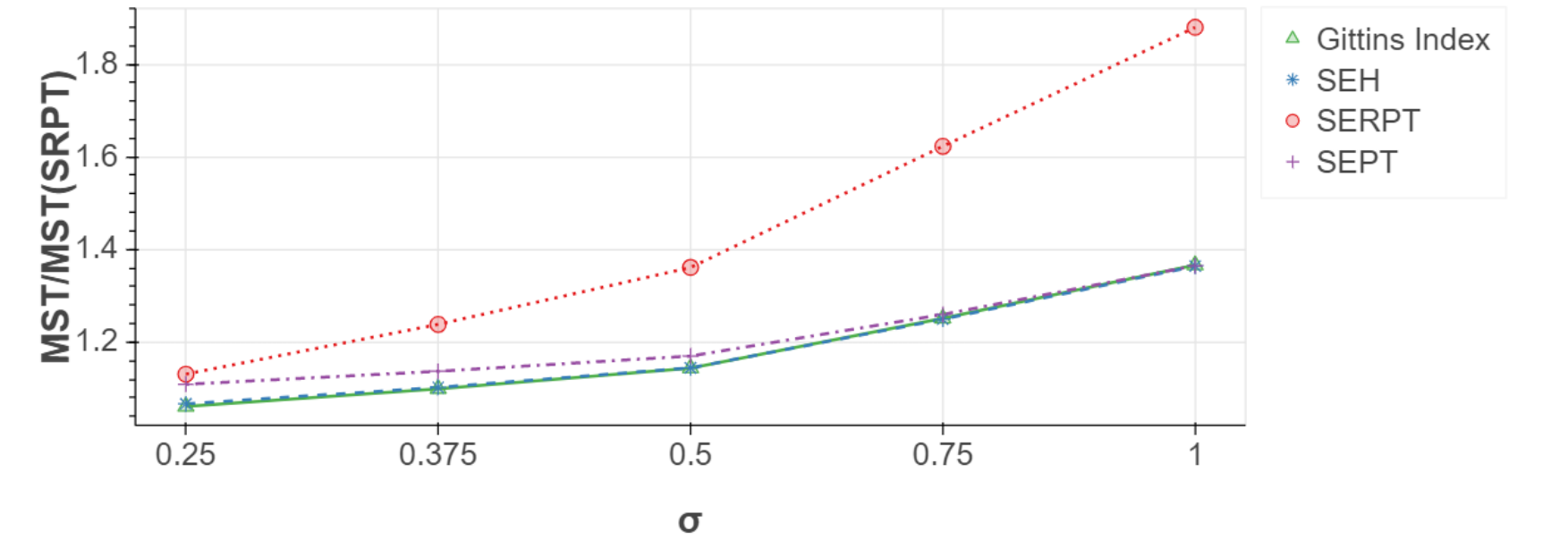}
  \centering
    \caption{MST of the Facebook Hadoop workload}
  \label{fig:HadoopMST}

\end{figure}

\textbf{Real Workloads} --- We consider a Facebook Hadoop cluster trace from $2010$ \cite{chen2012interactive} and show that the results with this workload look very similar to those with synthetic workloads generated with $ k = 0.25$. The trace consists of $24,443$ jobs. We assume each job's processing time is the sum of its input, intermediate output, and final output bytes. The job processing times of this workload have high variance, and thus, we run hundreds of simulations to reach the desired confidence interval (as described in Section \ref{SimulationParameters}). We vary the error estimation distribution's $\sigma$ parameter to evaluate different scenarios of estimated processing
time precision. To maintain the default settings in Table \ref{tab:1}, we define the processing speed in bytes per second. The arrival rate $\lambda$ is chosen to yield the desired $\rho = 0.9$. A simulation run ends when the last job in the workload arrives at the system and we calculate the MST of the jobs that are fully processed among the first $10,000$ jobs that entered the system. Fig. \ref{fig:HadoopMST} shows the MST normalized against the optimal MST obtained with SRPT with varying $\sigma$ between $0.25$ and $1$. We observe that the Gittins' Index and SEH policies perform best across all values of $\sigma$.

\begin{figure}[ht]

  \includegraphics[width=100mm,scale=0.5]{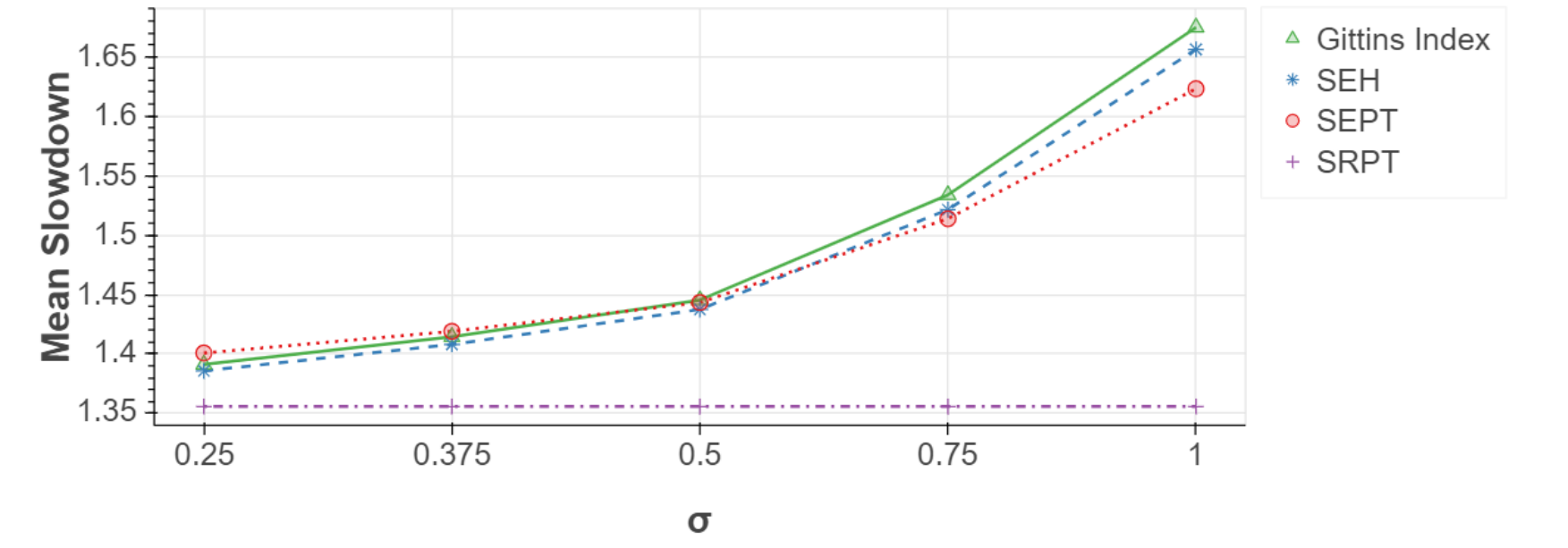}
  \centering
    \caption{Mean slowdown of the Facebook Hadoop workload}
  \label{fig:HadoopMSD}

\end{figure}

In Fig. \ref{fig:HadoopMSD}, we display the mean slowdown obtained with the policies under evaluation. Similar to Fig. \ref{fig:25SigmaMSD}, we have not included the mean slowdown of SERPT since it is several orders of magnitude higher. We observe that for $\sigma \le 0.5$, where the estimates are better, the SEH policy has lower mean slowdown than the Gittins' Index and SEPT policies, however, SEPT starts to outperform the Gittins' Index and SEH policies when $\sigma$ increases, consistent with our observations for synthetic workloads.

\section{Conclusion and Future Work} \label{Conclusion}

The SRPT policy, which is optimal for scheduling in single-server systems, may have problematic performance when job processing times are estimated. This work has considered the problem of scheduling with the presence of job processing time estimates. A multiplicative error model is used to produce estimation errors proportional to the job processing times. We have introduced a novel heuristic that combines the merits of SERPT and SEPT and requires minimal calculation overhead and no information about the job processing time and estimation error distributions. We have shown that this policy is consistent with a Gittins'-like view of the problem. Our numerical results demonstrate that the SEH policy has desirable performance in minimizing both the MST and mean slowdown of the system when there is low variance in the estimation error distribution. It outperforms SERPT except in scenarios where the job processing time variance is extremely low. Examining the SEH policy under other error models as well as analytic bounds as to how far it is from optimal could be investigated in future work. It would also be useful to examine how well policies designed for worst case performance would perform with respect to the performance metrics considered in this paper. The work of Purohit et al.\ \cite{purohit2018improving} is an intriguing candidate, as it runs two policies in parallel to provide worst case performance guarantees, even when there are large estimation errors.

Not much work has been done in the area of multi-server scheduling in the presence of estimation errors. One major reason is that determining optimal policies for multi-server queues is much more challenging compared to the single-server case. Mailach and Down \cite{mailach2017scheduling} suggest that when SRPT is used in a multi-server system, the estimation error affects the system's performance to a lesser degree than in a single-server system. Grosof et al.\ \cite{grosof2018srpt} prove that multi-server SRPT is asymptotically optimal when an M/G/$k$ system is heavily loaded. Our work only evaluates the performance of SEH in a single-server framework so we leave the extension and evaluation of this policy in multi-server queues for future investigation. \\
\\ \textbf{Acknowledgment}

The authors would like to thank Ziv Scully for useful discussions on the limitations of the SEH policy.
\bibliographystyle{splncs04}  
\bibliography{references}
\end{document}